ICSE - May 2018

# Fifty Years of Software Engineering

or

## The View from Garmisch

Brian Randell {Brian.Randell@ncl.ac.uk}

School of Computing

Newcastle University

**Abstract**:

On several earlier anniversaries of the 1968-69 NATO Software Engineering conferences I have acceded to requests to provide some reminiscences. I repeat some extracts from these reminiscences here, as a backdrop to brief comments on subsequent developments, and in particular on the distinctions between developing off-the-shelf package software and large one-off bespoke software systems. The software package industry had yet to come into existence in 1968-69, but has proved very successful. But some large software projects in the latter bespoke category still suffer from problems that are all too reminiscent of those that, in 1968, gave rise to discussion of a "software crisis".

On several earlier occasions, as anniversaries of the original 1968 and 1969 NATO Conferences on Software Engineering have loomed, I have accepted an invitation to reminisce on the original conferences and/or to comment on the subsequent progress of the subject. The first was on the 10th Anniversary of the conferences, at ICSE 1979, in Munich. This was in fact the first occasion after the 1969 conference that I had accepted any invitations to speak on software engineering, having been so disillusioned by the 1969 NATO conference that I had for years thereafter made a point of refusing to use the term or even be associated with any event that used it.

The paper that I was asked to provide for ICSE 1979 was one to be entitled "Software Engineering As It Was In 1968" [1]. Thus I concentrated on the first NATO Conference, which was held in Garmisch Partenkirchen , and ignored the follow-up 1969 conference on Software Engineering Techniques, which was held in Rome. Mine was one of four invited papers at ICSE 1979, the others being on "Software Engineering As It Is", by Barry Boehm [2], "Software Engineering As It Should Be", by Edsger Dijkstra [3] and "Software Engineering As It Will Be", by Wlad Turski [4].

Given how many years have since passed, I think it appropriate start by repeating the first part of my ICSE 1979:

Let me tell some of you, and remind others, about what the world of computing was like in the late 1960s. Computers were mainly very big (room size) and expensive, and were mainly used by rich organizations for large scale commercial data processing and scientific calculations, though the first mini-computers were starting to appear in well-funded laboratories.

The idea of a personal computer had not yet arrived. Data bases, networks and distributed systems were also yet to come, but work had started on implementing ARPANET, the Internet's main predecessor, though local area networks were unknown.

The term "software" came into use, though as yet systems software was usually provided "free" with the hardware by the computer manufacturer, and applications software was normally designed specially for particular environments and computers.

In those pre-Microsoft days, shifting to a new computer normally implied – because of hardware incompatibilities – users having to abandon or rewrite all existing applications programs.

Increasingly ambitious applications and systems software projects were being undertaken, and organizations found themselves becoming much more dependent on large and complex computer systems than had previously been the case. These various large software projects were almost all one-off projects, developed for specific customers, and all too many of the largest projects were characterized by "underestimates and overexpectations", as Licklider put it.

It was perhaps only when, in 1969, IBM "unbundled" its software by pricing it separately from its hardware that software became a commodity; and a recognisable software industry, and the notion of package software started to come into existence.

This then was the world of software that formed the background to the first NATO conference. In such circumstances, it is hardly surprising that the term "Software Engineering" was essentially unknown and unexpected in 1968, when it was chosen as a conference title by NATO's Study Group on Computer Science. I trust everyone understands that the title was originally chosen – I think by Fritz Bauer – as being *provocative*, describing a requirement rather than a reality.

In 1968 I was working at the IBM Research Center in Yorktown Heights, NY – and was invited by Fritz Bauer to join Peter Naur as co-editor of the planned conference report. (I should mention that the conference was aimed at producing a report that would be widely circulated to officials and governments, not just to the technical community.) Peter Naur was already famous as the Editor of the wonderful Algol 60 Report, and we were both members of the IFIP Working Group on Algol.

Many attendees have since described the Garmisch Conference as one of the most stimulating conferences that they have ever attended. A number of people, such as Edsger Dijkstra, were prompted or encouraged by the

> discussions to work on the problem of how to create programs that were mathematically-proven to be free from errors. In contrast, and appropriately reflecting my and Dijkstra's respective levels of programming skill, I became interested in the problem of how to design programs that could be usefully relied upon even if it was admitted that they still contained yet-to-be found bugs.
>
> In fact, a tremendously excited and enthusiastic atmosphere developed at the conference as participants came to realize the degree of common concern about what some were even willing to term the "*software crisis*", and general agreement arose about the importance of trying to convince not just other colleagues, but also policy makers at all levels, of the seriousness of the problems that were being discussed.

After ICSE 1979 I resumed my self-imposed embargo on writing about software engineering, in fact until 1996, when I was invited to Schloss Dagstuhl, for a conference that brought a group of computer historians together with a group of software practitioners to talk about the history of the subject. By this time I had taken up, and had published a book and a number of papers on, the subject of the history of digital computers, so I accepted the invitation to present a paper there, thinking of myself as a member of the first of these communities. However, I found I was being treated at Dagstuhl not so much as a computer historian as a historical artefact from the early software era!

The paper I gave at Dagstuhl in 1996 [5] concentrated on my involvement in the editing of the two software engineering reports – the 1968 one with Peter Naur [6], and the 1969 one with John Buxton [7]. It is symptomatic that my memories of Garmisch are mainly of the conference itself, of the fascinating technical discussions and excited atmosphere – whereas my memories of Rome are mainly of one wonderfully sarcastic talk, and of the post-conference trials and tribulations of producing a worthwhile report out of an otherwise largely unmemorable conference:

> [In fact the] NATO Committee decided there would be a follow-up conference, to be held in Rome in 1969. Peter Naur and I were again invited to edit the conference report. He wisely refused – I naively agreed, and it was arranged that I would be joined by John Buxton, then of Warwick University, one of the 1968 conference participants. . . .
>
> Unlike the first conference, at which it was fully accepted that the term software engineering expressed a need rather than a reality, in Rome there was a tendency on the part of some participants to talk as if the subject already existed. An even greater contrast to the Garmisch conference was the gulf that opened up between the researchers and practitioners, and the ill-tempered nature of many of the discussions.
>
> And it became clear during the conference that the organizers had a hidden agenda, namely that of persuading NATO to fund the setting up of an International Software Engineering Institute. However things did not go according to their plan. The discussion sessions, which were meant to provide evidence of strong and widespread support for this proposal, were instead marked by considerable scepticism, and led one of the participants, Tom

Simpson of IBM, to write and present a splendid short satire on "*Masterpiece Engineering*". The result of all this was to kill the idea of an International Software Engineering Institute stone dead.

John Buxton and I planned to include the text of "*Masterpiece Engineering*" in our report, as a concluding chapter, but were not allowed to do so. However, you can now find it on the web [8] – and I recommend to you do so – but meanwhile here is a small sample:

> Regrettably none of these advances in knowledge seemed to have any real impact on masterpiece production and so, at length, the group decided that the basic difficulty was clearly a management problem. One of the brighter students (by the name of L. da Vinci) was instantly promoted to manager of the project, putting him in charge of procuring paints, canvases and brushes for the rest of the organisation.

The comments I made at COMPSAC 2008 [9] some twelve years later, about subsequent software engineering developments, included the following:

> Since [the NATO conferences] not just one, but rather many, types of software industry have come into existence, in particular those that design or tailor "bespoke" software for particular clients and environments, and those that produce "off-the-peg" software packages that are sold to thousands or even millions of customers. The first type is a recognisable successor to the software activities of the 1960s. In the second, very different, type of software industry economies of scale, and Darwinian-style evolution, have a large impact on what sorts of software get implemented, and how such implementation is undertaken, e.g. involving getting hundreds or thousands of users involved, willingly or unwillingly, to help with software validation and refinement. Furthermore, this second type of industry is in the main concerned with what economists call a "natural monopoly", caused by the fact that though its development costs are immense, those required for manufacturing and distribution are comparatively trivial – so that being first to market, and achieving initial market dominance, are highly advantageous.
>
> Any reasonable account of how far we've come since the late 1960s (and where we've got to) has to treat these two types of software and software industry very differently. The first type of software industry has gone on to attempt ever larger and more complex tasks. But it is still subject to many of the same challenges concerning implementation cost, project schedule, performance and (especially) dependability that so exercised the NATO conference participants. The second now provides a wonderful marketplace of useable and useful software systems, utilities and applications that has utterly transformed society's utilisation and perception of computers. But technical monoculturalism, allied to the growth of computer networking, has led to this industry and its customers also suffering from all sorts of malicious, indeed criminal, activities that were not in any way foreseen in the discussions at the NATO conferences.

I must acknowledge that a much more expert, and more nuanced, view of software engineering is presented in my colleague Ian Sommerville's splendid text book on the subject, now into its tenth edition [10]. He argues that the subject really needs to be treated as covering a whole range of different sub-disciplines, relating to the differing environments, types of hardware system, and application domains. And the description and definition he gives of the overall subject area is as follows:

> Software engineering is an engineering discipline that is concerned with all aspects of software production from the early stages of system specification through to maintaining the system after it has gone into use.

I can contrast this view of *software engineering as an engineering discipline* with the much more critical assessment, in 2011, of the subject by David Parnas [11]:

> The most accurate title for the people that are now called "Software Engineers" would be "Paid Software Developers". They are professionals in the sense that they are paid for what they do, which is develop software but they have no credentials that are relevant to software development, the vast majority of the ones who are Professional Engineers gained that credential in another discipline and then migrated into software development without significant education about software design. . . When the term "Software Engineering" was introduced, many asked a simple question. "How is "Software Engineering different from programming"?

To my considerable surprise Parnas went on and said:

> The best answer to the questions posed above was provided by Brian Randell when he described Software Engineering as "The multi-person development of multi-version programs". This pithy phrase implies everything that differentiates software engineering from other programming.

My surprise is because I have no recollection of having coined such a phrase – and despite searching cannot find any evidence that would enable me to take the credit, much as I would like to. But what I also find surprising is that neither Sommerville nor Parnas draws such a harsh distinction as I did between bespoke and off-the-peg software, or places as much emphasis as I did, and do, on the significance of the Darwinian-style evolutionary forces operating in the software package marketplace.

The software systems that were such a matter of concern in 1968 were all large bespoke systems, typically designed ab initio and implemented for a single customer, and that customer's particular environment, and particular complex functional and dependability requirements. They were one-off projects, mostly aiming directly for very ambitious performance and dependability goals. Typically, they violated many of the apparently humorous but in fact deeply serious "rules" documented by John Gall in "*Systemantics*" [12] in particular:

> "A complex system that works is invariably found to have evolved from a simple system that worked."

As I have indicated, the contrast between large one-off bespoke projects and the software package marketplace could not be greater. Package designers typically first introduce relatively simple software systems to the marketplace, admittedly often somewhat prematurely, where these systems compete with others for customers (who often end up functioning, in their thousands, as unpaid system testers). Though advertising muscle can and does distort the marketplace, and in some arenas market domination makes it difficult for new entrants to gain a foothold, in many cases the effect is for a collective evolution in quality and functionality, and for survival of the fittest. Thus software can and often does emerge, eventually, that is of a quality and sophistication that far outstrips that which was such a cause for concern in Garmisch in 1968.

But meanwhile, the world of large one-off bespoke software systems continues to attempt straight off to build systems of great size, ambition and complexity, sometimes with results that seem all too similar to those discussed at Garmisch, despite the many claimed advances in software engineering tools and techniques.

One particular bespoke system, or rather system-of-systems, that comes to mind is one that the UK National Health Service's immense NPfIT Project tried to produce. This was claimed in 2005 to be the world's biggest civil IT project. Its statistics were staggering – the plan was to provide an online service for use by 40,000 general medical practitioners, 80,000 other doctors, 350,000 nurses, over 300 hospitals, 50 million patients, and well over a million healthcare workers throughout England! The facilities to be provided included detailed and summary electronic patient records, an electronic prescription service, radiographic image archiving and communication, appointment selection and booking, and a secure NHS email and personal demographics service.

I was one of a group of computer science professors who became very concerned about this project, and campaigned for it to be subjected to a stringent independent review. We wrote an open letter to the House of Commons Health Select Committee in April 2006, as a result of which we were invited to a meeting with the top management of the NPfIT Project. After a lengthy discussion both sides agreed that "a constructive and pragmatic independent review of the programme could be valuable" and that we would "meet again to consider further details of how such a review might best be conducted and its terms of reference". This agreement was publicised on the official NPfIT website and in the press – but then senior government officials stepped in and ruled out any such independent review, and the website new item was removed. And so the Project continued unchecked.

Over the next few years we created a very large online dossier of quotations from official reports, academic papers, newspaper accounts, etc., about the NPfIT Project and its problems [13]. My invited testimony to an enquiry into the Project by the House of Commons Health Select Committee drew heavily on the facts documented in this Dossier. This testimony was then the main basis for my subsequent publication "*A computer scientist's reactions to NPfIT*" [14].

In summary, our main concerns were (i) the huge size and complexity of the project, (ii) its over-centralised and very bureaucratic approach to system specification and acquisition, (iii) the inadequate consideration of socio-technical issues and pre-existing systems, and (iv) the extremely demanding, but inadequately specified, dependability requirements – in particular concerning safety, security and privacy. With regard to this last issue, the conclusion I came to was:

> "a very good summary of the fundamental security dilemma facing NPfIT is that one can (with difficulty) achieve any two of (a) high security, (b) sophisticated functionality, and (c) great scale – but achieving all three is currently (and may well remain) beyond the state of the art. . . . NPfIT looks

> set to sacrifice security; I believe that it should instead make every effort to evade the scale problem."

Needless to say, our fears proved well-founded, and the project was effectively disbanded in 2011, having met very few of its goals. It was subsequently described by the Government's powerful Public Accounts Committee as one of the "worst and most expensive contracting fiascos" in public sector history – it had a little earlier been admitted by the Government that the total cost was over £20 billion!

In 1968, at Garmisch, despite all the talk of a software crisis, it was recognized that many software projects were successful, and were producing satisfactory results. This surely is the case today, and a number of impressively large and sophisticated bespoke systems have been successfully implemented. But what almost forty years earlier Licklider had memorably termed "Underestimates and Overexpectations" [15] still occur – sometimes spectacularly so, as my brief account of NPfIT has indicated. (And of course NPfIT is not alone – see for example Computer Weekly's long-running blog on "*Public Sector IT – Exploring the challenges involved in large-scale IT projects in the public sector*" [16].)

Another concern I expressed ten years ago at COMPSAC about the world of software engineering was the multiplicity of "standards", something that led to software development being fragmented across a ridiculous number of different languages, methodologies, platforms, and standards. I quoted from Les Hatton's paper *Professionalism in IT [17]*.

> "We actively encourage students to work on real projects for industry and as a result one of my colleagues this year had students submitting projects in C, C#, C++, Java, PHP, MySQL, XML, HTML, XHTML, VB.Net on XP, Mac OS X, Linux and even Vista with Eclipse, Netbeans, Ant, JWSDP, Glassfish, DreamWeaver, Developer Studio, .Net with maybe even some Etruscan."

I find that the *Online Historical Encyclopaedia of Programming Languages* now lists no less than 8945 different programming languages [18]. This seems to me both highly excessive, and symptomatic of a rather worrying continuing situation, even though I fully accept Ian Sommerville's comment that, given the number and variety of sub-disciplines sheltering under the umbrella term "software engineering":

> "There are no universal software engineering methods and techniques that are suitable for all systems and all companies. Rather, a diverse set of software engineering methods and tools has evolved over the past 50 years."

This situation is understandable, but I am reluctant to accept that it justifies anywhere near 8945 languages, and the very large number of different methods and techniques that have been created. Rather I fear that the above criticisms by Parnas of the field retain much of their validity, now some seven years later, even though – as documented by, for example, Sommerville – the software engineering field has indeed made considerable progress.

I started my talk at ICSE 1979 with this quotation from Geoffrey Elton, the eminent Cambridge historian:

> "The future is dark, the present burdensome. Only the past, dead and finished, bears contemplation. Those who look upon it have survived it; they are its product and its victors. No wonder therefore that men concern themselves with history."

And another favourite quotation of mine is that by George Santayana [19]:

> "Those who cannot remember the past are condemned to repeat it".

So I hope that you will forgive me for having chosen to concentrate mainly on matters historical, and that these remarks will encourage at least some of you to pay a little more attention to the past. Indeed, let me urge you to take some time to read or re-read, for example, the original 1968 NATO Report, and what Barry Boehm, Wlad Turski and Edsger Dijkstra said of your subject back in 1979, before you resume work on inventing yet another new language or technique.

**References**

1. Randell, B.: 'Software Engineering: As It Was': '4th International Conference on Software Engineering. Munich, Germany' (IEEE Computer Society, 1979), pp. 1-10.

2. Boehm, B.W.: 'Software Engineering: As It Is': '4th International Conference on Software Engineering. Munich, Germany' (IEEE Computer Society, 1979), pp. 11-21.

3. Dijkstra, E.W.: 'Software Engineering: As It Should Be': '4th International Conference on Software Engineering' (IEEE Computer Society, 1979), pp. 442-448.

4. Turski, W.M.: 'Software Engineering: As It Will Be': '4th International Conference on Software Engineering. Munich, Germany' (IEEE Computer Society, 1979), pp. 449-456.

5. Randell, B.: 'The 1968/69 NATO Software Engineering Reports', IEEE Annals of the History of Computing, 1998, 20, (1), pp. 51-64.

6. Naur, P., and Randell, B.: 'Software Engineering: Report of a conference sponsored by the NATO Science Committee, Garmisch, Germany, 7th to 11th October 1968' (Scientific Affairs Division, NATO, 1969.)
[http://homepages.cs.ncl.ac.uk/brian.randell/NATO/nato1968.PDF (Accessed 16 April 2018)].

7. Buxton, J.N., and Randell, B.: 'Software Engineering Techniques: Report on a Conference sponsored by the NATO Science Committee, Rome, Italy, 27th to 31st October 1969' (Scientific Affairs Division, NATO, 1970.)
[http://homepages.cs.ncl.ac.uk/brian.randell/NATO/nato1969.PDF (Accessed 16 April 2018)].

8. Simpson, T.H.: 'Masterpiece Engineering'.
[http://homepages.cs.ncl.ac.uk/brian.randell/NATO/NATOReports/index.html#Appendix (Accessed 18 Apr 2018).

9. Randell, B.: 'Position Statement: How Far Have We Come': 'Proc. 32nd Annual IEEE International Computer Software and Applications Conference (COMPSAC)' (2008), p. 8.

10. Sommerville, I.: 'Software Engineering (10th ed.)' Pearson (2016) 810 pp. [ISBN 13:978-1-292-09613-1].

11. Parnas, D.L.: 'Software Engineering: Multi-person Development of Multi-version programs': 'Dependable and Historic Computing', eds. C.B. Jones and J.L. Lloyd (Springer 2011), pp. 413-427.

12. Gall, J.: 'Systemantics: How Systems Work & Especially How They Fail' (Times Books, 1977).

13. Anderson, R., et al: 'The NHS's National Programme for Information Technology (NPfIT): A Dossier of Concerns'. [http://homepages.cs.ncl.ac.uk/brian.randell/Concerns.pdf (Accessed 19 April 2018).]

14. Randell, B.: A computer scientist's reactions to NPfIT. Journal of Information Technology 2007, 22, (3), pp. 222-234.

15. Licklider, J.C.R.: 'Underestimates and Overexpectations': 'ABM: An Evaluation of the Decision to Deploy an Antiballistic Missile System' (Harper and Row, 1969), pp. 118-129.

16. Computer Weekly: 'Public Sector IT – Exploring the challenges involved in large-scale IT projects in the public sector'. [https://www.computerweekly.com/blog/Public-Sector-IT. (Accessed 18 Apr 2018)].

17. Hatton, L.: 'Professionalism in IT', Safety Critical Systems Club, 2008, 17, (2). [http://www.leshatton.org/Documents/Professionalism_2007.pdf (Accessed 16 April 2018)].

18. Pigott, D.: 'Online Historical Encyclopaedia of Programming Languages'. [http://hopl.info (Accessed 10 April 2018)].

19. Santayana, G.: 'The Life of Reason: Reason in Common Sense'. Dover Publications Inc., (New ed. of 1905 edition, 1980) 205 pp.